\documentclass[prd,superscriptaddress,showpacs,showkeys,twocolumn]{revtex4}
\usepackage[T1]{fontenc}
\usepackage{amsmath}
\usepackage{amssymb}
\usepackage[colorinlistoftodos]{todonotes}
\usepackage{epsfig}
\usepackage{palatino}
\def\be{\begin{equation}}
\def\ee{\end{equation}}
\usepackage[colorlinks=true,
            linkcolor=blue,
           urlcolor=blue,
           citecolor=blue]{hyperref}
\usepackage{tabstackengine}
\usepackage{graphicx}
\usepackage[english]{babel}
\usepackage{amsfonts}
\usepackage{eurosym}
\usepackage{calrsfs}
\usepackage{color}
\begin{document}

\title{Gravitational deflection of relativistic massive particles  by Kerr black holes and Teo wormholes viewed as a topological effect}

\author{Kimet Jusufi}
\email{kimet.jusufi@unite.edu.mk}
\affiliation{Physics Department, State University of Tetovo, Ilinden Street nn, 1200,
Tetovo, Macedonia.}
\affiliation{Institute of Physics, Faculty of Natural Sciences and Mathematics, Ss. Cyril and Methodius University, Arhimedova 3, 1000 Skopje, Macedonia.}

\begin{abstract}
We consider the problem of gravitational deflection of a propagating relativistic massive particles by rotating black holes (Kerr black holes) and rotating wormholes (Teo wormholes) in the weak limit approximation. In particular we have introduced an alternative way  to calculate the deflection angle for massive particles based on the refractive index of the optical media and the  Gauss-Bonnet theorem applied to the isotropic optical metrics. The refractive index governing the propagation of massive particles is calculated  by considering those particles as a de Broglie wave packets. Finally applying the Gauss-Bonnet theorem leads to an exact result for the deflection angle in both geometries.  Put in other words,  the trajectory of light rays as well as the trajectory of massive particles in a given spacetime background can be viewed as a global spacetime effect, namely as a topological effect. 

\end{abstract}


\date{\today}

\maketitle

\section{Introduction}
General theory of relativity provides an elegant mathematical relation of the spacetime geometry in one hand, and matter described by the energy-momentum tensor on the other hand. This theory explains many astrophysical phenomena and has been tested many times in the past. The experimental results strongly suggest an amazing agreement with theoretical predictions. Black holes and wormholes are mathematically predicted by general relativity. A traversable wormhole is a hyperspace tunnel, connecting together two distant regions within our universe or different univereses \cite{Wheeler55,Thorne88,Visser95}. Nowadays there is a lot of interesting research towards the quantum description of these objects in terms of quantum entanglement.

The bending of light is one of the most famous classical experiments confirming the curved nature of the spacetime geometry, furthermore this phenomenon is well explained in almost all the text books covering general relativity.  The standard explanation of light deflection is very simple; due to the presence of a massive body the light ray is deflected with an angle of deflection proportional to the mass of the system enclosed within a certain region usually known as the impact parameter. 

Gibbons and Werner, discovered yet another method offering a different perspective on the deflection of light. Namely, it shows the importance of topology on the trajectory of light rays in presence of a static and sperically symetric  gravitational field \cite{gibbons1}. This method involves the application of Gauss-Bonnet theorem (GBT)  in order to compute the deflection angle. In this approach, the deflection angle can be viewed as global effect, or to put in physics words, by integrating on a domain outside the light ray. A step forward was made by Werner who extended this method for asymtotically flat stationary metrics such as the Kerr black hole \cite{werner}. This approach is more complicated and involves the use of Finsler-Randers geometry. Furthermore, one has to apply the Naz{\i}m's method to construct a Riemannian manifold osculating the Randers manifold \cite{werner}. It is interesting to not that Werner's method has been extended to asymptotically non-flat statonary fields such as a presence of topological defects, namely a rotating cosmic string and a rotating global monopole \cite{kimet1,kimet2,kimet3,kimet4,kimet5}. Note that GBT was applied also to the strong limit and a finite distance corrections in the presence of the cosmological constant \cite{a1,a2,a3}.

Another important question which naturally arises is whether one can calculate the gravitational  deflection of massive particles using the GBT. As we shall see indeed this is the case. Very recently, Crisnejo and Gallo \cite{gallo} were able to compute the deflection angle for massive particles in a static spacetime geometry. 

Our main motivation in this paper is to address and solve the problem of computing the deflection angle for massive particles in a rotating spacetime geometry in the weak limit. Towards this purpose, we shall consider an alternative way to compute the deflection of light and massive particles in a Kerr black hole spacetime and Teo wormhole spacetime. In doing so, we will use an isotropic type metrics for a linearized rotating gravitational field which drastically simplifies the problem. Such a procedure for instance was applied in Refs. \cite{roy,nandi1,nandi2,paul} where authors investigated the deflection of light in terms of another method known as the optical-mechanical analogy.   The importance of isotropic metric relies in the fact that one can easily find the refractive index of the corresponding optical media. In order to study the case of massive particles we shall consider those particles as a de Broglie wave packets introduced in Ref. \cite{nandi1}. In fact, this will be crucial since a modification of the refractive index in a stationary gravitational field is needed.

This paper is organized as follows. In Section II we use an isotropic  metric form for the Kerr spacetime, after which we derive the refractive index of the Kerr optical media. We also compute the deflection angle for light rays using the GBT. In Section III, we consider the problem of the gravitational deflection of massive particles. In Section IV, we study deflection of massive particles in a Teo wormhole geometry. In Section V, we comment on our results. In this paper we use natural unites $G=c=\hbar=1$.

\section{Deflection of light in Kerr spacetime}

\subsection{The refractive index }
Let us start by writing the Kerr solution in the Boyer
Lindquist form given by
\begin{equation}\notag
ds^2=-\left(1-\frac{2M r}{\Sigma^2}\right)dt^2+\frac{\Sigma^2}{\Delta}dr^2+\Sigma^2 d\theta^2+\sin^2\theta d\varphi^2 
\end{equation}
\begin{equation}
\times \left(r^2+a^2+\frac{2Mr a^2}{\Sigma^2}\sin^2 \theta \right)-\frac{4Mr a \sin^2 \theta }{\Sigma^2}d\varphi dt
\end{equation}
with 
\begin{equation}
\Sigma^2=r^2+a^2 \cos^2\theta,
\end{equation}
\begin{equation}
\Delta=r^2-2Mr+a^2.
\end{equation}

To simplify the problem we can consider the deflection in the equatorial plane in a linearized Kerr metric in $a$. Such a metric can also describe the gravitational field around a rotating star or planet given by \cite{roy}
\begin{equation}
ds^2 \simeq -\left(1-\frac{2M}{r}+\frac{4Ma \frac{d\varphi}{dt} }{r}\right)dt^2+\frac{dr^2}{1-\frac{2M}{r}}+r^2d\varphi^2.
\end{equation}

Use the following coordinate transformation 
\begin{equation}
r=\rho \left(1+\frac{M}{2 \rho}\right)^2,
\end{equation}
after which the metric takes the following form \cite{roy}
\begin{equation}
ds^2=-\frac{(1-\frac{M}{2\rho})^2+\frac{4Ma}{\rho}\frac{d\varphi}{dt}}{(1+\frac{M}{2\rho})^2}dt^2+\left(1+\frac{M}{2\rho}\right)^4\left(d\rho^2+\rho^2 d\varphi^2\right).
\end{equation}

The last equation represents an isotropic metric, so
that the line element can be written in the form
\begin{equation}
ds^2=-\mathcal{F}^2(\rho) dt^2+\mathcal{G}^2(\rho) |d\vec{\rho}|^2.
\end{equation}

The isotropic coordinate speed of light $v(\rho)$ can be found from the relation
\begin{equation}
v(\rho)=|\frac{d\vec{\rho}}{dt}|=\frac{\sqrt{(1-\frac{M}{2\rho})^2+\frac{4Ma}{\rho}\frac{d\varphi}{dt}}}{(1+\frac{M}{2\rho})^3}.
\end{equation}

Using $n(\rho)=c/v(\rho)$ with $[c=1]$ the last equation yields the effective refractive index for light in the Kerr gravitational field 
\begin{equation}
n(\rho)=\frac{(1+\frac{M}{2\rho})^3}{\sqrt{(1-\frac{M}{2\rho})^2+\frac{4Ma}{\rho}\frac{d\varphi}{dt}}},
\end{equation}

In this way the optical metric reads
\begin{equation}
dt^2=n(\rho)^2 d\rho^2+\rho^2 n(\rho)^2 d\varphi^2.
\end{equation}

Note that the expression $d \varphi /dt$ can be computed by using the relativistic action function $S$ and the Hamilton-Jacobi equation. Without going into details one can show that the following relation holds 
\begin{equation}
\frac{d \varphi}{dt}=\frac{2M r a  E+(\Sigma^2-2M r)J}{\Sigma^2(r^2 E-2M r a J]}
\end{equation}
where $\Sigma^2=r^2$ in the equatorial plane. Note that $E$ is the energy of the particle (photon) given by $E=p c$ the angular momentum $J=p\, r_0$, $r_0$ being the distance of the closest approach which can be approximated with the impact factor, i.e. $r_0=b$ in the weak limit.  Using the definition  for the impact factor
\begin{equation}
\frac{J}{E}=b,
\end{equation}
we find
\begin{equation}
\frac{d \varphi}{dt}=\frac{2M a+(r-2M) b}{r^3-2Ma b}.
\end{equation}

By inverting the coordinate relation (5) one finds
\begin{equation}
\rho = \frac{1}{2}\left(r-M+\sqrt{r(r-2M)}\right),
\end{equation} 
which suggests that in leading order terms 
\begin{equation}
 \rho \simeq r-M +\mathcal{O}(M^2).
 \end{equation}

Hereinafter, we shall use the approximation $\rho \simeq r-M$. In particular we find the following relation for the refractive index 
\begin{equation}
n(r)=1+\frac{2M}{r}-\frac{2M a b}{r^3}+\mathcal{O}(M^2,a^2).
\end{equation}

This result clearly indicates that the refractive index is modified due to the angular momentum parameter $a$. 

\subsection{The GBT theorem and deflection of light}
Let us continue by rewriting the optical metric (10) in terms of a new coordinates. Namely introducing
\begin{eqnarray}
dr^{\star}&=&n(\rho)\, d\rho,\\
f(r^{\star})&=& n(\rho)\, \rho.
\end{eqnarray}

Now the optical metric reads
\begin{eqnarray}
\mathrm{d}t^{2}=\tilde{g}_{ab}\,\mathrm{d}x^{a}\mathrm{d}x^{b}=\mathrm{d}{%
r^{\star }}^{2}+f^{2}(r^{\star })\mathrm{d}\varphi ^{2}. 
\end{eqnarray}
 
In terms of the coordinates the Gaussian optical curvature $\mathcal{K}$ is expressed as: 
\begin{eqnarray}
  \mathcal{K} & = & - \frac{1}{f (r^{\star})}  \frac{\mathrm{d}^2 f
  (r^{\star})}{\mathrm{d} r^{\star 2}} \\\notag
  & = & - \frac{1}{f (r^{\star})}  \left[ \frac{\mathrm{d} \rho}{\mathrm{d}
  r^{\star}}  \frac{\mathrm{d}}{\mathrm{d} \rho} \left( \frac{\mathrm{d}
  \rho}{\mathrm{d} r^{\star}} \right) \frac{\mathrm{d} f}{\mathrm{d} \rho} + \left(
  \frac{\mathrm{d} \rho}{\mathrm{d} r^{\star}} \right)^2 \frac{\mathrm{d}^2
  f}{\mathrm{d} \rho^2} \right]. 
\end{eqnarray}

Furthermore we can express the last equation in terms of the refraction index resulting with

\begin{align}\label{Curvature}
\begin{split}
\mathcal{K} = & -\frac{n(\rho) n''(\rho) \rho-(n'(\rho))^2\rho+n(\rho) n'(\rho)}{n^4(\rho) \rho}
\end{split}.
\end{align} 

With these results in hand, we can chose a non-singular region $\mathcal{D}_{R}$ with boundary $\partial
\mathcal{D}_{R}=\gamma _{\tilde{g}}\cup C_{R}$. The GBT provides a connection between geometry (in a sense of optical curvature) and topology (in a sense of Euler characteristic number) stated as follows
\begin{equation}
\iint\limits_{\mathcal{D}_{R}}\mathcal{K}\,\mathrm{d}S+\oint\limits_{\partial \mathcal{%
D}_{R}}\kappa \,\mathrm{d}t+\sum_{i}\theta _{i}=2\pi \chi (\mathcal{D}_{R}),
\end{equation}
with $\kappa $ being the geodesic curvature, and $\mathcal{K}$  being the Gaussian optical curvature. Following \cite{gibbons1} we can choose a non-singular domain  with Euler characteristic number $%
\chi (\mathcal{D}_{R})=1$. The geodesic curvature is defined 
\begin{equation}
\kappa =\tilde{g}\,\left(\nabla _{\dot{%
\gamma}}\dot{\gamma},\ddot{\gamma}\right),
\end{equation}
but also keeping in mind an additional unit speed condition $\tilde{g}(\dot{\gamma},\dot{%
\gamma})=1$, with $\ddot{\gamma}$ being the unit acceleration vector. The two corresponding jump angles in the limit $R\rightarrow \infty $, reads  $\theta _{\mathit{O}%
}+\theta _{\mathit{S}}\rightarrow \pi $. Therefore the GBT now is simplified as follows
\begin{equation}
\iint\limits_{\mathcal{D}_{R}}\mathcal{K}\,\mathrm{d}S+\oint\limits_{C_{R}}\kappa \,%
\mathrm{d}t\overset{{R\rightarrow \infty }}{=}\iint\limits_{\mathcal{D}%
_{\infty }}\mathcal{K}\,\mathrm{d}S+\int\limits_{0}^{\pi +\hat{\alpha}}\mathrm{d}\varphi
=\pi.
\end{equation}

The geodesic curvature $\kappa$ can be easily computed. Taking into consideration that $\kappa (\gamma _{\tilde{g}})=0$ (remember $\gamma _{\tilde{g}}$ is a
geodesic), we are left with the following contribution
\begin{equation}
\kappa (C_{R})=|\nabla _{\dot{C}_{R}}\dot{C}_{R}|.
\end{equation}

Without loss of generality, in the large radial limit we can set $C_{R}:=r(\varphi)=R=\text{const}$, implying the radial component relation
\begin{equation}
\left( \nabla _{\dot{C}_{R}}\dot{C}_{R}\right) ^{r}=\dot{C}_{R}^{\varphi
}\,\left( \partial _{\varphi }\dot{C}_{R}^{r}\right) +\tilde{\Gamma} _{\varphi
\varphi }^{r}\left( \dot{C}_{R}^{\varphi }\right) ^{2}. 
\end{equation}

A direct computation reveals that
\begin{eqnarray}\notag
\lim_{R\rightarrow \infty }\kappa (C_{R}) &=&\lim_{R\rightarrow \infty
}\left\vert \nabla _{\dot{C}_{R}}\dot{C}_{R}\right\vert , \notag \\
&\rightarrow &\frac{1}{R},
\end{eqnarray}%
with a similar simplification from the optical metric
\begin{eqnarray}
\lim_{R\rightarrow \infty } \mathrm{d}t&\to & R \, \mathrm{d}\varphi.  
\end{eqnarray}%

If we combine the last two equations, we find that in fact our optical geometry is asymptotically Euclidean
\begin{equation}
\lim_{R \to \infty}\frac{\kappa (C_{R})\mathrm{d}t}{\mathrm{d}\,\varphi}= 1
\end{equation}

The deflection angle then is found by solving the integral 
\begin{eqnarray}
\hat{\alpha}=-\int\limits_{0}^{\pi}\int\limits_{r_{\gamma}}^{\infty} \mathcal{K}  \sqrt{\tilde{g}} \,\,dr^{\star} d\varphi.
\end{eqnarray}

Using (16) and (20) for the Gaussian optical curvature we find
\begin{equation}
\mathcal{K} \simeq -\frac{2M}{r^2}+\frac{18 \,a M b }{r^5}+\mathcal{O}(M^2,a^2).
\end{equation}

It is important to note that the total deflection angle measured by a static observer at infinity consists of $\hat{\alpha}=\hat{\alpha}_{M}+{\hat{\alpha}}_{Ma}-\delta$, where $\hat{\alpha}_{M}$ is the deflection angle proportional to the black hole mass, ${\hat{\alpha}}_{Ma}$ is a contribution coming from the rotating spacetime, while $\delta$ must be considered due to the induced frame dragging effect as a result of the co-rotating metric (4). In this way, $\delta$ must be taken into account in order to compute the deflection angle measured by a static observer at infinity.  So intuitively, one expects the following relation $\delta = {\hat{\alpha}}_{Ma}/2$, resulting with $\hat{\alpha}=\hat{\alpha}_{M}+{\hat{\alpha}}_{Ma}/2$, corresponding to the static observer at infinity.  Namely for the prograde case, the deflection angle for such an observer at infinity results with
\begin{eqnarray}
\hat{\alpha}=-\int\limits_{0}^{\pi}\int\limits_{\frac{b}{\sin \varphi}}^{\infty} \left(-\frac{2M}{r^2}+\frac{1}{2}\frac{18 \,a M b }{r^5} \right) \sqrt{\tilde{g}} \,\,dr^{\star} d\varphi,
\end{eqnarray}
where  we have used the light ray equation 
\begin{equation}
r=\frac{b}{\sin \varphi}.
\end{equation}

Note that we have also used
\begin{equation}
dS=\sqrt{\tilde g}\, dr^{\star} d\varphi=n(\rho)^2 \rho d\rho d\varphi \simeq r dr d\varphi,
\end{equation}

Solving the last integral we find the total deflection angle 
\begin{equation}
\hat{\alpha}\simeq \frac{4M}{b} \pm \frac{4Ma}{b^2}.
\end{equation} 

This is exactly the same result found by Werner using the Rander-Kerr optical geometry \cite{werner}. We also note that the signs of positive and negative stand for a retrograde and a prograde light rays, respectively. In particular the retrograde case follows from the symmetry $a \to -a$.

\section{Deflection of massive particles in Kerr spacetime}
\subsection{Index of refraction for massive-particle de Broglie waves}
Recall from  geometrical optics that the refractive index can be used to find the trajectory of light by varying the path between two fixed points in space 
\begin{equation}
\delta \int_{x_1}^{x_2} n dl=0
\end{equation}
where $dl=|d\vec{r}|$ is the element of the path of
integration in the three-dimensional space. The orbits of a relativistic particles one the other hand are obtained by requiring that they be geodesics: 
\begin{equation}
\delta \int_{x_1,t_1}^{x_2,t_2} ds=0
\end{equation}

Alternatively, the last equation can be written in accordance with the Hamilton's principle as follows
\begin{equation}
\delta \int_{t_1}^{t_2} L(x_i,w_i) dt=0,
\end{equation}
where the componentes of the three relativistic velocity are given by $w_i=dx_i/dt$ and $w^2=\sum_i^3 w_i^2$. The effective Lagrangian is written as \cite{nandi1}
\begin{equation}
L(x_i,w)=-m  \mathcal{F} \sqrt{1-w^2 n^2}
\end{equation}

The Hamiltonian  is found to be
\begin{equation}
H=m  \mathcal{F}\left(1-w^2 n^2\right)^{-1/2}.
\end{equation}

In Ref. \cite{nandi1} authors have shown that from Hamilton's principle one can derive an analogous relation to Fermat's principle written as
\begin{equation}
\delta \int_{x_1}^{x_2} p dl=\delta \int_{x_1}^{x_2} (H n^2 w) dl=0.
\end{equation}

Furthermore, using $p=\hbar k$ and $H=E=\hbar \omega $, (note that we have temporarily introduced $\hbar$) leads to the following de Broglie wavelength of a given massive particle
\begin{equation}
\lambda=\frac{h c}{n H \sqrt{1-\frac{m^2 c^4 \mathcal{F}^2 }{H^2}}}.
\end{equation}

This equations that can be easily rearranged as
\begin{equation}
\lambda n \sqrt{1-\frac{m^2 c^4 \mathcal{F}^2 }{H^2}}=\frac{h c}{ H }=const.
\end{equation}

In other words, this equation is a generalisation  of a  well known result in wave-optics. For massive particles, in a given isotropic metric, the expression should be constant everywhere in the optical medium, i.e. 
 $\lambda N=const$. The refractive index relation for massive particles in that case reads
\begin{equation}
N(r)=n(r)\sqrt{1-\frac{m^2}{E^2}\mathcal{F}^2(r)},
\end{equation}
 
We can apply this expression to our Kerr optical media which results with
\begin{equation}
\mathcal{F}^2(r)=\frac{(1-\frac{M}{2 r})^2+\frac{4Ma}{r}\frac{d\varphi}{dt}}{(1+\frac{M}{2 r})^2},
\end{equation}
in which the approximation $\rho \simeq r$ is used. One can write the above relation in leading order terms 
\begin{equation}
\mathcal{F}^2(r)=1-\frac{2M}{r}+\frac{4\, a \,M \,b\, w}{r^3}+\mathcal{O}(M^2,a^2).
\end{equation}

On the other hand the particles energy measured at infinity is given by
\begin{equation}
E =\frac{m}{\left(1-w^2\right)^{1/2}},
\end{equation}
in which $m$ is the rest mass of the particle and $w$ is the relativistic velocity. The angular momentum can be written as
\begin{equation}
J=\frac{m w b}{\left(1-w^2\right)^{1/2}},
\end{equation}
where $b$ is the impact parameter. Following the definition of the impact parameter we can write
\begin{equation}
\frac{J}{E }=w\,b,
\end{equation}
which reduces to $b$ in the case of light $w=c=1$. Therefore, the quantity  $d \varphi /dt$  in the case of massive particles is modified as follows
\begin{equation}
\frac{d \varphi}{dt}=\frac{2M a+(r-2M) w b}{r^3-2Ma w b},
\end{equation}
yielding
\begin{equation}
N(r)=w+\frac{w M(r^2 w^2-2 a b w+r^2)}{w^2 r^3}+\mathcal{O}(M^2,a^2).
\end{equation}

\subsection{Gravitational Deflection angle of massive-particles }

We can, therefore, express the Gaussian optical curvature for massive particles in terms of the modified refractive index as follows
\begin{align}
\begin{split}
\mathcal{K} = & -\frac{N(\rho) N''(\rho) \rho-(N'(\rho))^2\rho+N(\rho) N'(\rho)}{N^4(\rho) \rho}.
\end{split}
\end{align} 

In particular, in leading order terms we find
\begin{equation}
\mathcal{K} \simeq  -\frac{M(r^2w^2-18 a b w+r^2)}{r^5 w^4}+\mathcal{O}(M^2,a^2).
\end{equation}

Furthermore we can apply exactly the same procedure as in the case of light deflection. For instance the geodesic curvature results with
\begin{eqnarray}\notag
\lim_{R\rightarrow \infty }\kappa (C_{R}) &=&\lim_{R\rightarrow \infty
}\left\vert \nabla _{\dot{C}_{R}}\dot{C}_{R}\right\vert , \notag \\
&\rightarrow &\frac{1}{w R}. 
\end{eqnarray}%

For an observer located at a very large distance we can also write 
\begin{eqnarray}
\lim_{R\rightarrow \infty } dt&\to & w R \, d\varphi.  
\end{eqnarray}%

Again, in order to find the deflection angle measured by a static observer at infinity we use  $\delta = {\hat{\alpha}}_{Ma}/2$, resulting with $\hat{\alpha}=\hat{\alpha}_{M}+{\hat{\alpha}}_{Ma}/2$. Thus from the GBT it follows the  expression 
\begin{equation}
\hat{\alpha}=-\int\limits_{0}^{\pi}\int\limits_{\frac{b}{\sin \varphi}}^{\infty}\left(-\frac{M(r^2w^2+r^2)}{r^5 w^4}+\frac{1}{2}\frac{18 a M b }{r^5 w^3}\right)\sqrt{\tilde{g}} \,dr^{\star} d\varphi.
\end{equation}

We can approximate the surface element in terms of $r$ as follows
\begin{equation}
dS=\sqrt{\tilde g} \,dr^{\star} d\varphi=N(\rho)^2 \rho d\rho d\varphi \simeq w^2 r dr d\varphi,
\end{equation}
This integral can easily be evaluated, yielding
\begin{equation}
\hat{\alpha}\simeq \frac{2M}{b}\left(1+\frac{1}{w^2}\right)\pm \frac{4Ma}{b^2}\frac{1}{w}.
\end{equation}

Interestingly, we recovered the gravitational deflection angle of massive particles in a Kerr geometry which is in perfect agreement with the result reported in Ref. \cite{rr1,rr2}. Moreover in the special case, $w=c=1$ the light deflection angle is recovered.

\section{Deflection of massive particles by Teo wormhole spacetime}

The Teo wormhole metric represents a stationary wormhole solution given by the following metric \cite{teo}
\begin{equation} 
ds^2=-N^2 dt^2+\frac{dr^2}{\left(1-\frac{b_0}{r}\right)}+r^2 K^2\left[d\theta^2+\sin^2 \theta \left(d\varphi-\omega dt\right)^2\right],
\end{equation}
where
\begin{eqnarray}
N&=&K=1+\frac{\left(4 a \cos\theta\right)^2}{r},\\
\omega &=&\frac{2a}{r^3}.
\end{eqnarray}

Where $a$ is referred to the spin angular momentum, $b_0$ represents the shape function with the conditions $r \geq b_0$. The throat of the wormhole is located at the coordinate $r=b_0$. The flare-out condition reads \cite{naoki}
\begin{equation}
\frac{b_0-b_{0,r}r}{2 b_0^2}>0.
\end{equation}

As in the Kerr spacetime, we simplify the problem by considering a linearized rotating wormhole in $a$. We, therefore,  in the equatorial plane find
\begin{equation}
ds^2 \simeq -\left(1+\frac{4a }{r} \frac{d\varphi}{dt}\right)dt^2+\frac{dr^2}{1-\frac{b_0}{r}}+r^2d\varphi^2.
\end{equation}

Introducing a further coordinate transformation 
\begin{equation}
r=\rho \left(1+\frac{b_0}{4 \rho}\right)^2,
\end{equation}
yielding an isotropic Teo wormhole metric
\begin{equation}
ds^2=-\frac{(1+\frac{b_0}{4\rho})^2+\frac{4a}{\rho}\frac{d\varphi}{dt}}{(1+\frac{b_0}{4\rho})^2}dt^2+\left(1+\frac{b_0}{4\rho}\right)^4\left(d\rho^2+\rho^2 d\varphi^2\right).
\end{equation}

Without going into details, the expression $d \varphi /dt$  is computed as follows
\begin{equation}
\frac{d \varphi}{dt}=\frac{2 a +b w r}{ r^3-2 a b w}
\end{equation}

With this result in hand, in the case of the wormhole optical media we find the following result for the refractive index
\begin{equation}
N(r)=w+\frac{b_0 w}{2r}-\frac{2 a b w}{r^3}+\mathcal{O}(b_0^2,a^2).
\end{equation}

Utilizing the expression for the Gaussian optical curvature and the the refractive index we find
\begin{align}
\begin{split}
\mathcal{K} \simeq & -\frac{b_0}{2 r^3 w^2}-\frac{18 ab}{r^5 w^3}+\mathcal{O}(a^2,b_0^2)
\end{split}
\end{align} 

Setting $w=c=1$ we recover the case massless case. The geodesic curvature is modified as
\begin{eqnarray}\notag
\lim_{R\rightarrow \infty }\kappa (C_{R}) &=&\lim_{R\rightarrow \infty
}\left\vert \nabla _{\dot{C}_{R}}\dot{C}_{R}\right\vert , \notag \\
&\rightarrow &\frac{1}{w R},
\end{eqnarray}%
together with the relation
\begin{eqnarray}
\lim_{R\rightarrow \infty } dt&\to & w R \, d\varphi.  
\end{eqnarray}%

This implies  $
\kappa (C_{R})\mathrm{d}t= \mathrm{d}\,\varphi
$. Finally  is we substitute  the Gaussian optical curvature and applying the relation $\hat{\alpha}=\hat{\alpha}_{M}+{\hat{\alpha}}_{Ma}/2$, the gravitational deflection angle measured by a static observer at infinity we find the following relation
\begin{equation}
\hat{\alpha}=-\int\limits_{0}^{\pi}\int\limits_{\frac{b}{\sin \varphi}}^{\infty}\left( -\frac{b_0}{2 r^3 w^2}-\frac{1}{2}\frac{18 ab}{r^5 w^3} \right)\sqrt{\tilde{g}} \,dr^{\star} \,d\varphi.
\end{equation}

Evaluating the above integral we find
\begin{equation}
\hat{\alpha}\simeq \frac{b_0}{b} \pm \frac{4 a}{b^2 w}.
\end{equation}

In the special case, letting $w=c=1$, the above result reduces to the deflection angle of light reported in \cite{kimet7}. It is interesting to observe that the geometric contribution to the deflection angle remains invariant by the nature the particles.  Similarly, the signs of positive and negative stand for a retrograde and a prograde light rays, respectively.

\section{Conclusions}

In the present paper we have studied the gravitational deflection of massive and massless particles in a linearized Kerr and Teo spacetime backgrounds.  We have introduced a new approach to compute the gravitational deflection angle based on the refractive index and the GBT applied to an isotropic type metrics. 
The importance of this method relies in the fact that one can compute the deflection angle of relativistic particles in terms of the refractive index $N(r)$ by assuming the propagating massive particles as a de Broglie wave packets, resulting with an important constant quantity $\lambda N=const$, in a given optical media.
This is different, say, to the Werner's approach which involves the use of Finsler geometry. The refractive index governing the propagating of massive particles in is found by considering those particles as a de Broglie wave packets. 

Consequently, we have shown that the refractive index for massive particles is affected by the angular momentum parameter $a$, mass of the black hole, the wormhole shape function $b_0$, and finally the relativistic velocity of the particle $w$. Applying the GBT to the isotropic metrics we have found the following results for the gravitational deflection angles:

\begin{equation}\notag
\hat{\alpha}_{Kerr}\simeq \frac{2M}{b}\left(1+\frac{1}{w^2}\right)\pm \frac{4Ma}{b^2}\frac{1}{w},
\end{equation}
in the case of Kerr spacetine confirming \cite{rr1,rr2}, and
\begin{equation}\notag
\hat{\alpha}_{Teo}\simeq \frac{b_0}{b} \pm \frac{4 a}{b^2 w}
\end{equation}
in the case of Teo wormhole spacetime. It is interesting to point out that the geometric contribution to the deflection angle in the wormhole geometry  remains invariant and not affected by the mass of the particles i.e. $b_0/b$. In contrast, the spin angular contribution as well as the mass of the black hole is affected by the relativistic velocity of the particle $w$. 
From the last two equations we observe that an apparent singularity appears when $w\to 0$, therefore an additional constraint should be imposed, namely $0 < w \leq 1$. In principle this apparent singularity can be resolved, for instance there should be some critical value $w_{min}<w$ which solves the problem of this apparent singularity. In a very recent article \cite{sing} authors discuss such a critical value in the framework of the geodesic approach  for the charged black hole. Moreover they find $w_{min}^2<(M/b)^{3/2}$ which removes the singularity only for the Schwarzschild case, however there is no satisfactory solution for a more general case. It will be interesting to see if the deflection angle of a slowly moving particles can be found in terms of the GBT.

Finally our results clearly suggest the importance of the spacetime topology on the lensing effect, namely the gravitational deflection of massive particles can be viewed as a global effect. Lensing of particles might be an important tool in astrophysics in order to distinguish black holes from wormholes by their deflection angles of massive particles, such as the lensing of massive neutrinos.  We plan to extend this method by adding addition fields, such as scalar and electromagnetic fields. In addition it will be also interesting to study finite distance corrections on the deflection angle where the source and observer are located at a finite distance from each other.

\end{document}